\documentclass[a4paper]{article}
\usepackage[fontsize=12bp]{fontsize}
\usepackage[margin=2cm]{geometry}
\usepackage[utf8]{inputenc}
\usepackage[T1]{fontenc}
\usepackage{graphicx}
\usepackage{amsmath,amsfonts}
\usepackage{siunitx}
\usepackage{xcolor}
\usepackage{chemist}
\usepackage{multirow}
\usepackage{indentfirst}
\usepackage{float}
\usepackage[pdftex,colorlinks=true,linkcolor=blue,citecolor=blue,urlcolor=blue]{hyperref}
\RequirePackage{pdflscape}

\usepackage{siunitx}

\begin{document}
\begin{titlepage}
       \centering
       \huge
       \vspace*{1cm}
       \textbf{Predicting RNA-small molecule binding sites by 3D structure}

    \begin{center}        
       \large
       \vspace*{7cm}
       \textbf{Author : PAN Nan \\[1cm]{Supervisor : WALDISPÜHL Jérôme (McGill University)}}

       \vspace*{3cm}

       McGill University
       \vspace*{3cm}
       
       From January to June 2023
       \vspace*{3cm}
       
\centering
\includegraphics[width=5cm]{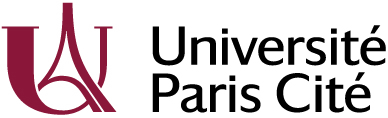}~
\includegraphics[width=5cm]{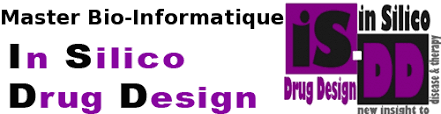}~
\includegraphics[width=6cm]{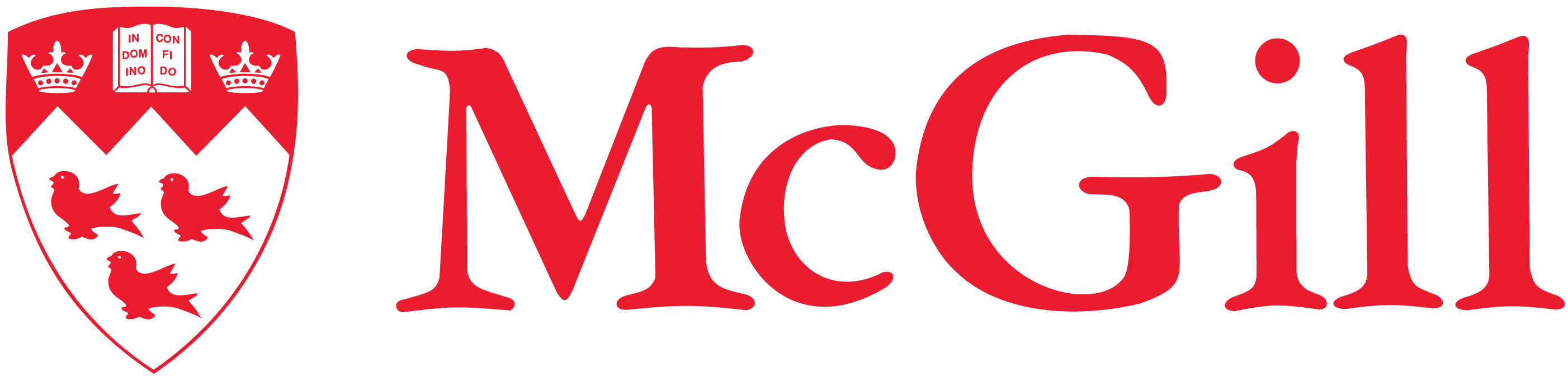}~

\end{center} 
\end{titlepage}

\thispagestyle{empty}

\textbf{Acknowledgment}\\

I would like to express my gratitude to my supervisor, Jérôme Waldispühl. His inspiration led me to incorporate RNA secondary structure types into my machine learning models. I am also thankful for his strong recommendation to other professors when I was seeking a PhD position.

\newpage

\thispagestyle{empty}

\textbf{Preface}\\

This internship has indeed been an exceptionally unique experience for me. I was granted an international internship scholarship for this opportunity. However, I wasn't able to travel to Montreal due to my visa being issued towards the latter part of my internship period. Consequently, I have been working remotely and unfortunately do not have access to my scholarship and internship salary.

Nonetheless, this circumstance did not impede my work. In fact, starting from Christmas 2022, I had been deeply engrossed in devising strategies to achieve the prediction of RNA-small molecule binding sites and formulating an optimal algorithm to attain this objective.

The internship started on 16 January 2023 and defended on 20 June 2023. The initial version of the algorithm was completed a week before the internship began. While it wasn't flawless, it provided sufficient evidence to reinforce my belief that my approach could indeed enable the prediction of RNA-small molecule binding sites. Consequently, I presented my concepts and the initial algorithm results during the group meeting on the inaugural day of the internship. Subsequently, I had the liberty to pursue my own ideas throughout the course of the internship.

Approximately two-thirds of the internship and the primary tasks are carried out in Geneva. This includes defining research ideas, designing the algorithm, composing the internship report, and preparing for the internship defence. And the remaining take place in Paris.

The report is not good enough because I did it independently in a limited amount of time without any professional help. And due to certain reasons unrelated to scientific research that I personally cannot compromise on, the results of this report might not be publishable as a scientific article. 

However, research is around discovery and dissemination. Personally, what holds the utmost significance for me is the research process itself – one that I relish, entailing the resolution of each puzzle and the conquering of every challenge. Hence, despite the hindrance in publishing this outcome, my inclination remains strong to share it with you. Research stands as an unending pursuit, and together, let's revel in the journey!

Should my research pique your curiosity, feel free to reach out to me at any time!

\newpage

\thispagestyle{empty}

\textbf{Abstract}\\

The prediction of RNA-small molecule binding sites is crucial for the discovery of effective drugs. Various computational methods have been developed to address this challenge, using information about the structure and sequence of RNA. In this study, we introduce \textit{CplxCavity}, a combination of a new algorithm and a machine learning model specifically designed to predict RNA-small molecule binding sites. \textit{CplxCavity} leverages the 3D structure of RNA or RNA complexes to identify surface cavities that have the potential to bind with small molecules. Our results demonstrate that \textit{CplxCavity} outperforms existing methods by accurately identifying binding sites for small molecules on RNA or RNA complexes. The introduction of \textit{CplxCavity} represents a significant advancement in computational tools for studying RNA-ligand interactions, and offers promising prospects for accelerating drug discovery and the development of therapies targeting RNA.

\vspace*{4cm}

\textbf{Résumé}\\

La prédiction des sites de liaison ARN-petites molécules est cruciale pour la découverte de médicaments efficaces. Diverses méthodes informatiques ont été développées pour relever ce défi, en utilisant des informations sur la structure et la séquence de l'ARN. Dans cette étude, nous introduisons \textit{CplxCavity}, une combinaison d'un nouvel algorithme et d'un modèle machine learning spécialement conçu pour prédire les sites de liaison entre l'ARN et les petites molécules. \textit{CplxCavity} exploite la structure 3D de l'ARN ou du complexe d'ARN pour identifier les cavités de surface qui présentent un potentiel de liaison avec les petites molécules. Nos résultats démontrent que \textit{CplxCavity} surpasse les méthodes existantes en identifiant avec précision les sites de liaison des petites molécules sur l'ARN ou les complexes d'ARN. L'introduction de \textit{CplxCavity} représente une avancée significative dans les outils informatiques pour l'étude des interactions ARN-ligand et offre des perspectives prometteuses pour accélérer la découverte de médicaments et le développement de thérapies ciblant l'ARN.

\newpage
\tableofcontents
\thispagestyle{empty}

\newpage
\setcounter{page}{1}

\section{Introduction}

RNA is a highly versatile molecule that performs various essential functions within the cell, such as catalysing biochemical reactions, serving as a structural component of ribosomes, and regulating gene expression~\cite{Mattick2006,Wang2022}. Small molecules are organic compounds capable of binding to RNA molecules and modulating their function. This opens up potential therapeutic applications in medicine. Small molecules that target RNA show great promise as drugs due to their high specificity, potency, and ability to target RNA molecules that are often unreachable by conventional small molecule drugs~\cite{Childs2022,Connelly2016,Matthew2019,Yu2020}.

The identification of binding sites between RNA or RNA complex and small molecules is a crucial step in the process of drug discovery and development. Various experimental techniques, such as X-ray crystallography~\cite{Jackson2023,Turnbull2021}, NMR spectroscopy~\cite{Sun2019}, SAXS (small angle X-ray scattering)~\cite{Magbanua2014}, and fluorescence spectroscopy~\cite{Magbanua2014,Bernacchi2007,Huranova2009}, have been employed to identify these binding sites and study the interactions between RNA or RNA complex and ligand.

Recent advances in computational methods have allowed for the prediction of RNA-small molecule interactions and the design of novel drugs. Molecular docking, molecular dynamics simulations, and machine learning algorithms are among the most commonly used techniques for this purpose. 

There are several existing computational methods for predicting RNA binding sites. Rsite and Rsite2 respectively measure the Euclidean distance between each nucleotide and all other nucleotides in the RNA tertiary structure or RNA secondary structure~\cite{Zeng2015, Zeng2016}. RBind uses a structural network approach~\cite{WangK2018}. RNAsite predicts the RNA binding sites by using the sequence and structure-based descriptors~\cite{Su2021}. Their method performs rather well on a small benchmark dataset, however, the performance is not guaranteed to be stable for a larger dataset.

In this context, we introduce a new approach called \textit{CplxCavity} for predicting RNA-small molecule binding sites. This approach uses a two-step process to predict the binding sites based on the 3D structure. Firstly, \textit{CplxCavity} algorithm directly determines and extracts the cavities on the surface of the structure by using the atomic coordinates of the residues located on the structure's surface. Secondly, we use a machine learning model to predict whether the identified cavity is an approximate binding site or non-binding site. 

Comparing these tools, RNAsite stands out as the most recent tool and has demonstrated superior performance compared to others~\cite{Su2021}. Therefore, we are comparing our tool with RNAsite. The first distinction is that our benchmark dataset is larger than theirs. Our dataset contains 330 RNA-small molecules complexes. In RNAsite's benchmark dataset, there are two groups: RB19, which comprises 19 RNA complexes, and RB78, which consists of 78 RNA complexes. Furthermore, our own dataset already includes 11 RNA complexes from RB19 and 38 RNA complexes from RB78. The remaining complexes that are not part of our dataset introduce the second difference, which is the definition of a ligand. In this study, we define a ligand as a small molecule with a molecular weight of less than $\SI{1000}{daltons}$~\cite{Macielag2012}. Examples of small molecules include natural metabolites and other drug-like compounds. In other studies, metal ions and solvent molecules are also considered as ligands~\cite{Zeng2015, Zeng2016, WangK2018, Su2021}. However, for the purpose of RNA-small molecule binding site research, they are not relevant. The objective is to examine the interaction between small molecules and RNA or RNA complexes in order to enhance the drug discovery process, rather than merely identifying locations where binding can occur.

RNAsite uses MCC (\textit{Matthews correlation coefficient})~\cite{Chicco2020,Chicco2021} and AUC (\textit{Area under the ROC Curve})~\cite{Hanley1982,FAWCETT2006} to assess its performance. In the RB78 group, RNAsite achieves an MCC of 0.186 and an AUC of 0.703. Comparatively, for the RB19 group, the MCC improves to 0.526, and the AUC increases to 0.834. The MCC ranges from -1 to 1, with a value of 0 indicating performance no better than random guessing. These results indicate that RNAsite may not be suitable for application in the RB78 group. It demonstrates promising performance in the RB19 group. But, the RB19 group comprises only 19 RNA complexes. In contrast, our tool, \textit{CplxCavity}, employs a dataset with 330 RNA-small molecule complexes. The MCC and AUC for the final model are 0.504 and 0.829, respectively. This implies that \textit{CplxCavity} outperforms RNAsite in terms of MCC and AUC metrics, and it can be applied to a wider variety of RNA complexes.

The current study is organized as follows. After providing the dataset in Section~\ref{sec2}, we introduce the \textit{CplxCavity} algorithm and the machine learning model associated with it, and evaluation metrics used for evaluating model performance. Section~\ref{sec3} presents the algorithm's results and the predicted results with detailed discussion. We conclude in Section~\ref{sec4} with some comments and remarks on our study.

\section{Materials and methods}\label{sec2}

We introduce our approach \textit{CplxCavity} for predicting RNA-small molecule binding sites. We first clarify the dataset used in this study, and then describe the components of the \textit{CplxCavity}.

\subsection{Dataset}
\paragraph{HARIBOSS dataset}

Our project uses the HARIBOSS (\textit{Harnessing RIBOnucleic acid-Small molecule Structures}) dataset~\cite{Panei2022}, which consists of 746 RNA-small molecule complexes in CIF (\textit{Crystallographic Information File}) format. These complexes have been determined using various methods such as X-ray crystallography, nuclear magnetic resonance spectroscopy, and cryo-electron microscopy. The dataset contains a variety of complexes, including RNA-small molecule, RNA-DNA-small molecule, and RNA-protein-small molecule complexes. The dataset comprises 1268 binding sites, which are defined as all the residues within $\SI{6}{\angstrom}$ of the ligand atoms, and RNA chains with more than 10 atoms. This dataset is an invaluable resource for studying the interactions between RNA and small molecules.

\paragraph{Interaction analysis}

The ligands in this study are defined as small molecules with a molecular weight of less than $\SI{1000}{daltons}$~\cite{Macielag2012}. Examples of such small molecules include natural metabolites and other drug-like compounds. We study all the complexes of HARIBOSS using PLIP 2021 (\textit{Protein–ligand interaction profiler 2021})~\cite{Adasme2021} to verify the existence of interaction between the RNA or RNA complex and the ligand. PLIP 2021 is an updated version of the PLIP software to analyze and determine the interactions between ligands and various biomolecules, including DNA, RNA, and proteins. The output comprises a collection of identified interactions, encompassing seven distinct types of interactions, hydrogen bond, hydrophobic contact, stacking-$\pi$, cation-$\pi$ interaction, salt bridge, water bridge, and halogen bond~\cite{Salentin2015}. Regrettably, PLIP 2021 cannot be used to analyze all the complexes in HARIBOSS due to format limitations. Indeed, PLIP 2021 is designed to process files in the PDB (\textit{Protein Data Bank}) format, but 178 of the complexes in HARIBOSS do not have the corresponding PDB files. Since large structures are often not available in PDB format, they cannot be analyzed using PLIP 2021.

\paragraph{Data filtering}
The PLIP 2021 algorithm considers seven types of interactions, and among all, the cation-$\pi$ interaction is the longest, with a distance of $\SI{6}{\angstrom}$. This finding is supported by~\cite{Justin1999}. In order to obtain a relatively plausible and stable binding site, we define the binding site as the set of residues within $\SI{6}{\angstrom}$ of the ligand that satisfy specific criteria : the number of residues should be greater than 3, the number of interacting residues should be greater than 2, and the number of amino acids should be less than one-third of the total number of residues. 

\paragraph{Benchmark dataset}
By following these criteria, we create a benchmark dataset comprising 330 RNA-small molecule complexes. Moreover, we identify 571 binding sites within these complexes. The size of these 571 binding sites varies from 4 to 25 residues, with an average of 13 residues. These binding sites are primarily composed of nucleic acids, although they may also include amino acids.

\subsection{\textit{CplxCavity}}

\begin{figure}[htpb]
    \centering
    \includegraphics[width=15cm]{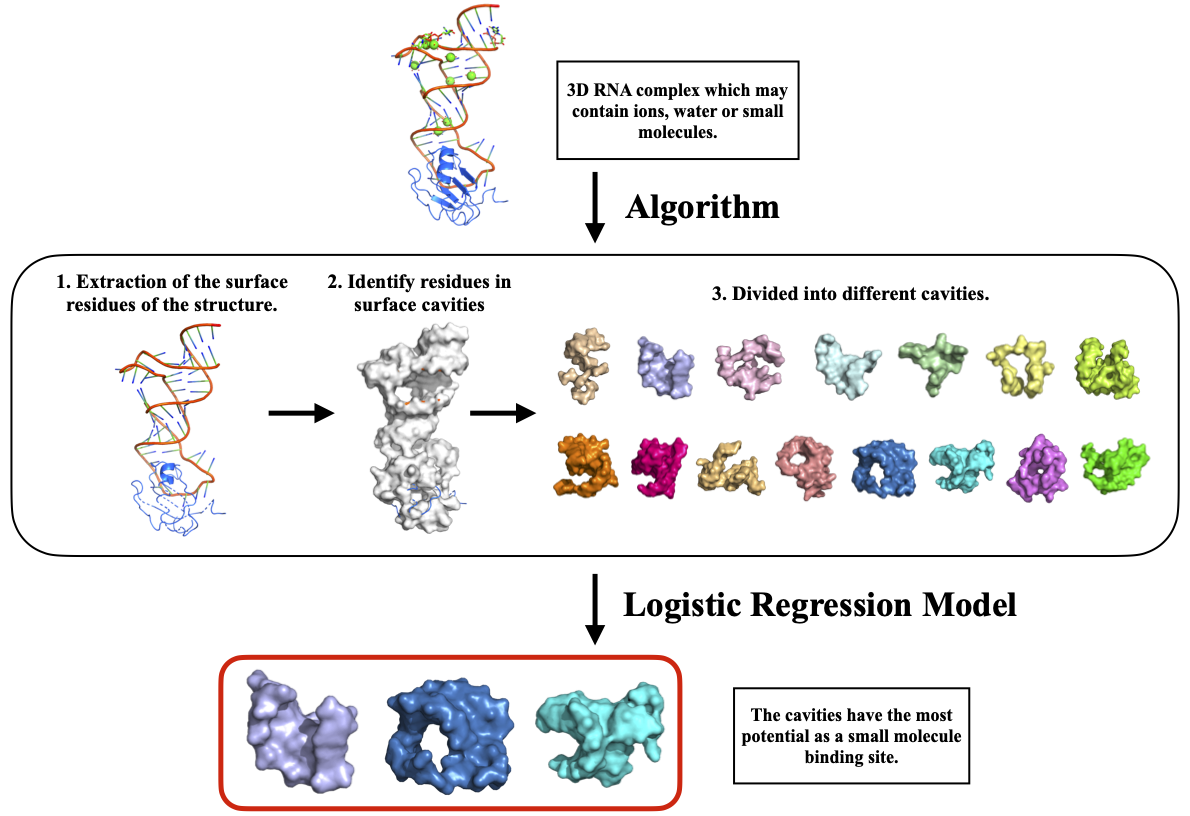}
    \caption{Representation of \textit{CplxCavity} strategy.}
    \label{Fig.strategy}
\end{figure}

The aim of \textit{CplxCavity} is to identify binding sites on the surface of RNA or RNA complexes from PDB file that may contain ions, water, and small molecules. Our approach consists of two main steps: cavity detection and classification. Figure~\ref{Fig.strategy} presents the strategy of \textit{CplxCavity}. We first use the algorithm to determine the cavities on the surface of the structure, and then use a trained logistic regression model to predict whether the cavities are approximate binding sites or non-binding sites.

\subsubsection{Algorithm}
We construct an algorithm to determine all the cavities on the surface of the structure. To achieve this, we first use the PyMOL~\cite{PyMOL} module in Python to extract the surface of the structure. We then use k-means clustering~\cite{MacQueen1967} to group surface residues based on the 3D coordinates of their atoms. To determine the optimal number of groups for using k-means clustering, we use the minimum number of residues of the binding sites in the benchmark dataset as the threshold, which is 4. The number of groups is then defined by $k := \frac{n}{4}$, where $n$ is the total number of surface residues. The value of $k$ can influence the sensitivity of the algorithm. 
Let us consider the group $g^i$ with $i\in [1,k]$. For two elements $a,b\in g^i$, we define the matrix for each group $g^i$ by 
\[\begin{aligned}
M_{g^i_{a,b}}&=\left\{
\begin{aligned}
&0, \quad&& \text{if}~ \lVert (a-O_{g^i}),(b-O_{g^i}) \rVert = 0 ~\text{and}~ \lVert b \rVert < \lVert a \rVert, \\
& &&\text{or}~ \frac{\lVert (a-O_{g^i}),(b-O_{g^i}) \rVert}{\lVert a \rVert \lVert b \rVert} = 1 ~\text{and}~ \lVert b \rVert \leq \lVert a \rVert,\\
& &&\text{or}~ \lVert (a-O_{g^i}),(b-O_{g^i}) \rVert = -1, \\
& &&\text{or}~ \left|\frac{\lVert a, O_{g^i} \rVert}{\lVert (a-O_{g^i}),(b-O_{g^i}) \rVert}\right| > 1,\\
&1, \quad&& \text{if}~ \lVert (a-O_{g^i}),(b-O_{g^i}) \rVert = 0 ~\text{and}~ \lVert b \rVert \geq \lVert a \rVert, \\
& &&\text{or}~ \frac{\lVert (a-O_{g^i}),(b-O_{g^i}) \rVert}{\lVert a \rVert \lVert b \rVert} = 1 ~\text{and}~ \lVert b \rVert > \lVert a \rVert,\\
&   && \text{or}~ \left|\frac{\lVert a, O_{g^i} \rVert}{\lVert (a-O_{g^i}),(b-O_{g^i}) \rVert}\right| \leq 1,\\
\end{aligned}
\right.
\end{aligned}\]
where $O_{g^i}$ is the centre of the group $g^i$. With the help of the matrix $M_{g^i}$, we then compute the score $S_{g^i_a}$ for each atom by
\begin{equation*}
    S_{g^i_a} = \frac{\sum_{b\in g^i} M_{g^i_{a,b}}}{m_{g^i}},
\end{equation*}
where $m_{g^i}$ represents the number of atoms for each group.

For all atoms in the group $g^i$, if the score of an atom is relatively greater than other atoms, this atom has higher probability of being in the cavity. For this reason, we focus on atoms which have a local maximum value on the score curve. However, not all atoms that present a local maximum value need to be studied, since the surface of the structure is not smooth. 

\begin{figure}[htpb]
    \addtolength{\leftskip} {-0.5cm}
    \includegraphics[width=18cm]{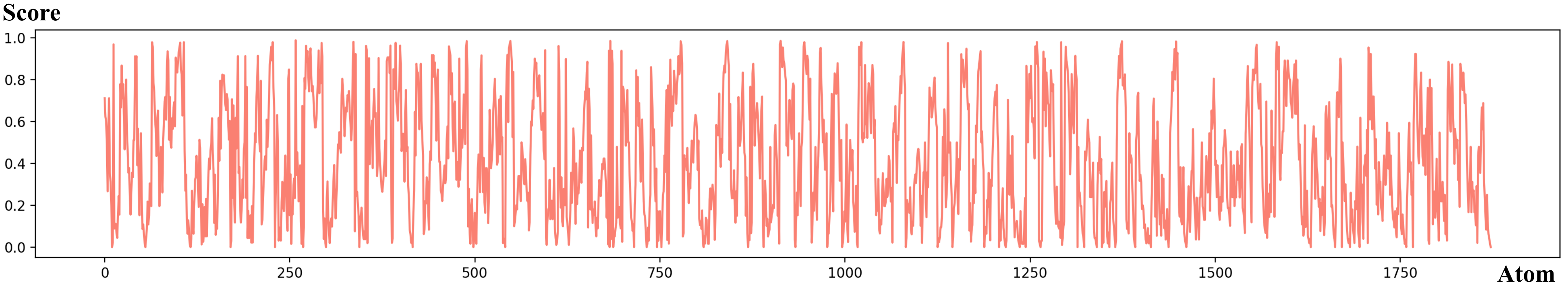}
    \caption{Score for each atom on the surface of PDB 7D7V.~\cite{Chen2020}.}
    \label{Fig.ex_score}
\end{figure}

An example of a score curve for the surface of PDB 7D7V~\cite{Chen2020} is presented in Figure~\ref{Fig.ex_score}. 
The score curve does not need to be smoothed, as it could reduce sensitivity. We identify atoms that have a local maximum value on the score curve, and the score value of the atom should be greater than the third quartile of the score curve (i.e., 80\% if the number of surface residues < 30). We then consider residues containing any of these atoms as part of the surface cavity, namely the residues in cavities. The number of the residues in cavities is denoted by $\theta$.

Based on the 3D coordinates of the atoms of residues in cavities, we group them into different cavities using k-means clustering. We use the average number of residues of the binding sites in the benchmark dataset, which is 13, to determine the optimal number of cavities. The number of cavities is defined by $\frac{\theta}{13}$. In the case of multiple cavities interconnected, we need to identify the residues that form the connecting part for each combination of cavities. We then can compute the geometric centre for each connecting part. This centre can be used as the centre of a new cavity. We consider all residues within $\SI{14}{\angstrom}$ from the centre to define the new cavity. This choice is based on the size of the small molecules being roughly between $\SI{10}{\angstrom}$ and $\SI{20}{\angstrom}$. We use the average of this interval as an estimate to define the size of the ligand, and use the distance of $\SI{6}{\angstrom}$ to define the binding site. If the set of residues differs from any previously determined cavities, it will be considered as a new cavity.

\begin{figure}[htpb]
    \centering
    \includegraphics[width=15cm]{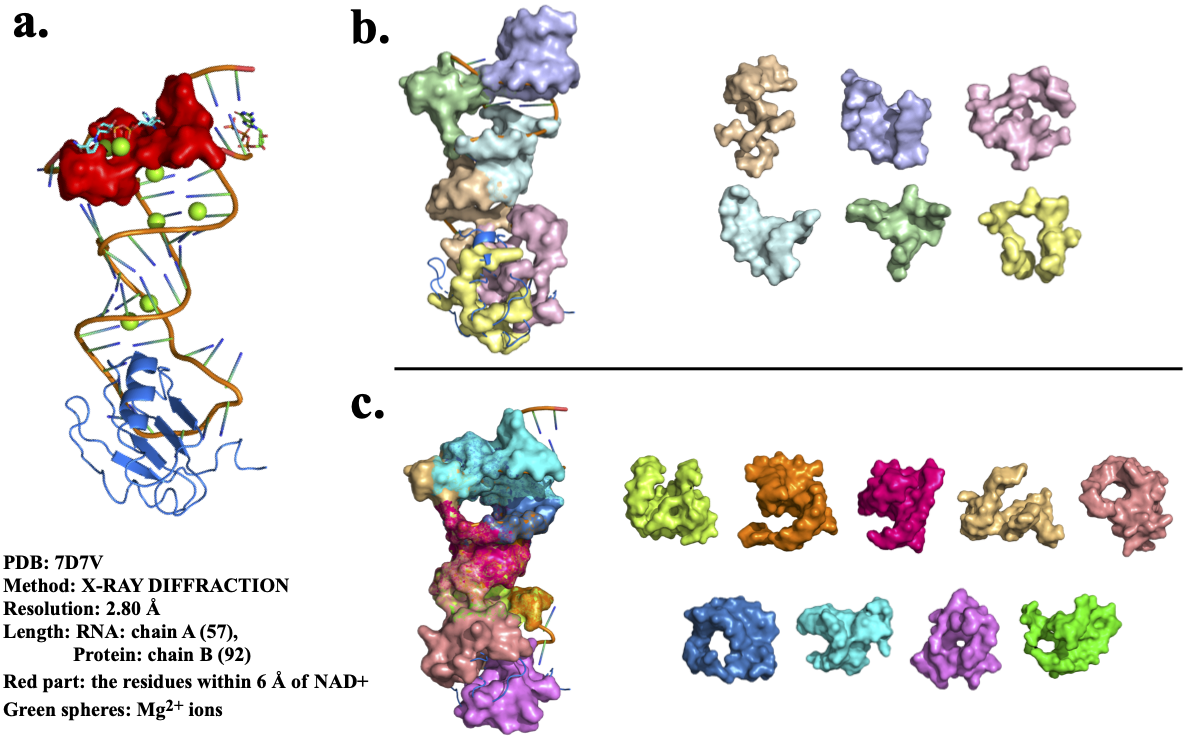}
    \caption{Identified cavities of PDB 7D7V~\cite{Chen2020} using our algorithm. a: The 3D structure of PDB 7D7V, the red part represents the binding site of the ligand NAD+. b: The cavities determined by k-means clustering. c: The cavities identified from the connecting parts of the connected cavities in b.}
    \label{Fig.ex_algo}
\end{figure}

Figure~\ref{Fig.ex_algo} shows an example of the determined cavities of PDB 7D7V~\cite{Chen2020} using our algorithm. 
Figure~\ref{Fig.ex_algo}a presents the 3D structure of PDB 7D7V, its structural information and the binding site (red part) of the ligand NAD+. There are six cavities determined by k-means clustering as shown in the Figure~\ref{Fig.ex_algo}b. Figure~\ref{Fig.ex_algo}c presents the nine cavities identified from the connecting parts of the connected cavities in the Figure~\ref{Fig.ex_algo}b.

\paragraph{Cavity analysis}

We observe in Figure~\ref{Fig.ex_algo}a and Figure~\ref{Fig.ex_algo}b that several cavities overlap with the binding site. We thus use PID (\textit{Percentage identity})~\cite{Karlin1990,Vogt1995,Raghava2006} to identify the approximate binding or non-binding sites from the cavities determined. Firstly, we remove all the cavities consisting only of amino acids. We then analyze the remaining cavities using PID,
\[\text{PID} = \frac{\text{Number of residues of (Binding site $\cap$ Cavity)}}{\text{Number of residues of Binding site}}.\]
If PID$\geq0.5$, we consider the cavity as an approximate binding site. Conversely, if PID$<0.5$, the cavity will be considered as an approximate non-binding sites.

\subsubsection{Machine learning}

After extracting cavities from PDB file, we require a model to determine whether the extracted cavity is an approximate binding site or non-binding site. We use the algorithm to study our benchmark dataset to generate a new dataset, this new dataset comprises 50\% approximate binding sites, while the remaining 50\% consists of approximate non-binding sites. We then analyze their characteristics through three distinct categories: geometry, physicochemistry, and RNA secondary structure type. This comprehensive assessment involves a total of 68 descriptors, as detailed in the Appendix.
For the analysis of nucleic acid structures, we use Barnaba~\cite{Bottaro2019}, a Python library. Additionally, we apply Forgi 2.0, another Python library, to study the RNA secondary structure~\cite{Thiel2019}. The computation of residue depth is performed using Biopython~\cite{cock2009}.
In order to build a classification model, 80\% of this dataset will be used as a training set to train the model, with the remaining part being the test set. In this study, we use the logistic regression model~\cite{Cox1958} to predict RNA-small molecule binding sites.

Then, we construct an initial model incorporating all the descriptors. Subsequently, we apply the AIC (\textit{Akaike’s information criterion})~\cite{Burnham1998} and analyze the p-values associated with the z-values of descriptors~\cite{Becker1988, Johnson1996,Venables2003} to simplify the model.
The z-value and the p-value associated with the z-value are defined by 
\[\text{z-value} = \frac{\text{Estimate}}{\text{Standard error}},\]
\[\text{p-value associated with the z-value} = 2 \times (1-\frac{1}{2}(\text{erf}(\frac{\text{z-value}}{\sqrt{2}}))),\]
where erf is the error function~\cite{Andrews1998}.
If the p-value associated with the z-value of the descriptor is less than 0.05, it suggests that the response descriptor has a statistically significant relationship with the predictor descriptor in the model. Consequently, we retain this descriptor in the model. This step helps streamline the model and enhances its effectiveness.

Next, we compare the performances between both of models using MCC (\textit{Matthews correlation coefficient})~\cite{Chicco2020,Chicco2021}, confusion matrix~\cite{STEHMAN1997}, and ROC curve (\textit{Receiver operating characteristic curve})~\cite{Hanley1982, FAWCETT2006} to determine the final model. 
When assessing binary classifications, MCC offers a more informative and accurate score. The value of MCC is between -1 and 1. In the case of perfect misclassification, a value of -1 is achieved. Conversely, in the case of perfect classification, a value of 1 is attained. An MCC value of 0 indicates that the performance of the classifier is equivalent to random guessing. 

From the confusion matrix, we can calculate various performance metrics, including accuracy, sensitivity, specificity, and PPV (\textit{positive predictive value}), which are represented as values between 0 and 1. A value of 1 indicates a perfect performance for the respective metric. The accuracy measures the overall correctness of the predictions. In this study, sensitivity and specificity are used to describe the accuracy of the model in predicting approximate binding sites and non-binding sites. PPV is the proportion of approximate binding sites among all predicted approximate binding sites.

AUC (\textit{Area under the ROC Curve}) is a measure that can be calculated from the ROC curve. It represents the area under the ROC curve and ranges between 0 and 1. A value of 1 indicates a perfect classifier, while a value of 0.5 suggests a random classifier.

\section{Results}\label{sec3}

\subsection{Algorithm}

We identify 3863 cavities from the 330 RNA-small molecule complexes in the dataset by using the algorithm of \textit{CplxCavity}. To evaluate the accuracy of our approach, we compare the 3863 cavities with their associated binding site by computing the PID. Finally, we determine 1383 approximate binding sites and 2480 approximate non-binding sites. 
\begin{figure}[h]
    \centering
    \includegraphics[width=10cm]{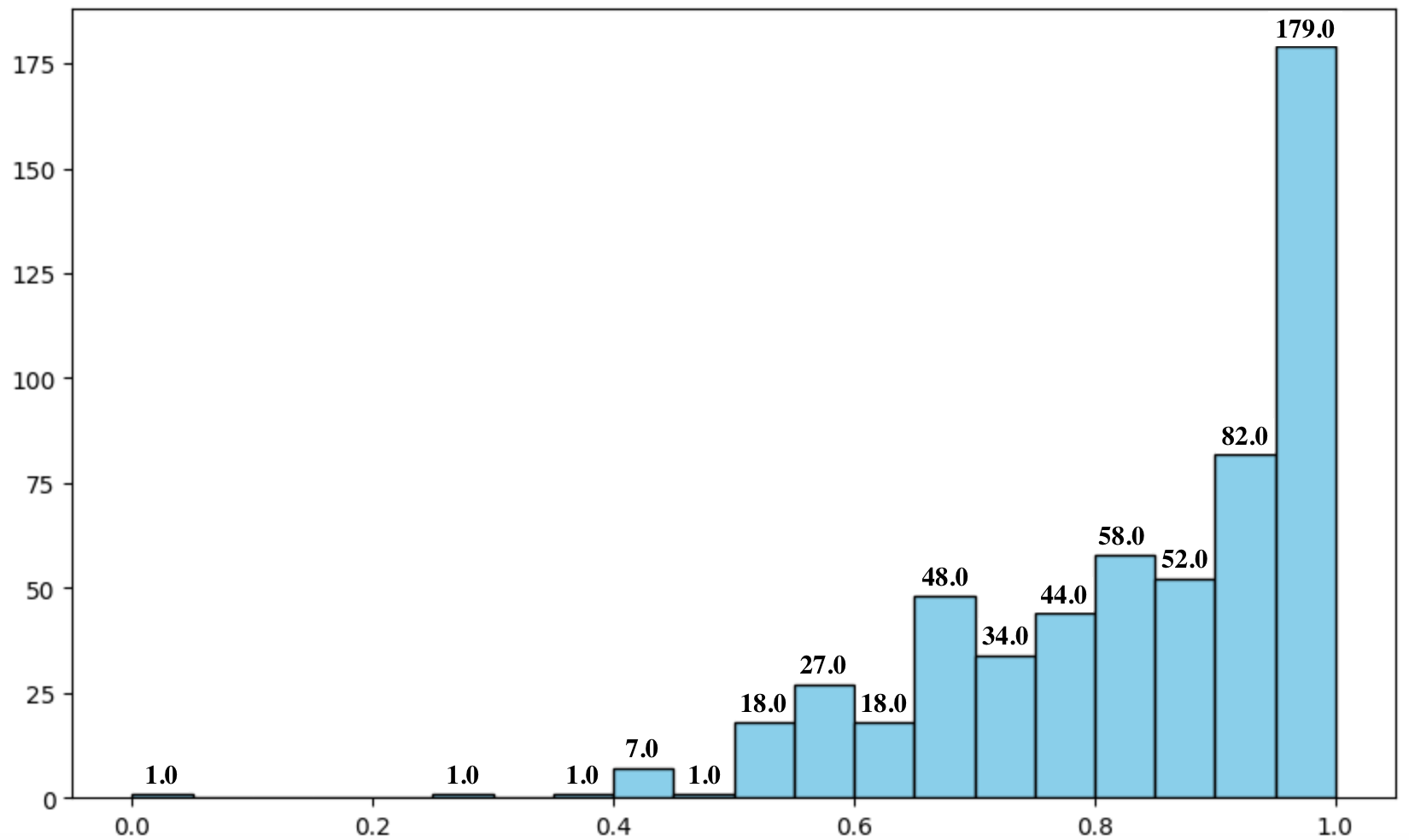}
    \caption{Distribution of the best PIDs (\textit{Percentage identity}) for the 571 binding sites.}
    \label{Fig.dist_algo}
\end{figure}

Figure~\ref{Fig.dist_algo} shows the distribution of the best PIDs for the 571 binding sites. We observe that there are 560 binding sites with the best PID greater than or equal to 0.5, indicating that 98.07\% of the binding sites are computed accurately. Meanwhile, 371 binding sites have an approximate binding site with a PID above 0.80. These results show that almost all the binding sites are contained in the cavities on the surface of the structure as determined by the algorithm, making the algorithm a reliable method for finding them.

\subsection{Machine learning model}

In order to construct a logistic regression model, we create a new dataset using 1383 approximate binding sites and 1383 approximate non-binding sites. The 1383 approximate non-binding sites are randomly selected from 2480 approximate non-binding sites. We then compute all 68 descriptors (see Appendix). Figure~\ref{Fig.corrplot} shows the correlation between descriptors after data cleaning. We observe that the 34 descriptors are independent of each other. 
\begin{figure}[htpb]
    \centering
    \includegraphics[width=10cm]{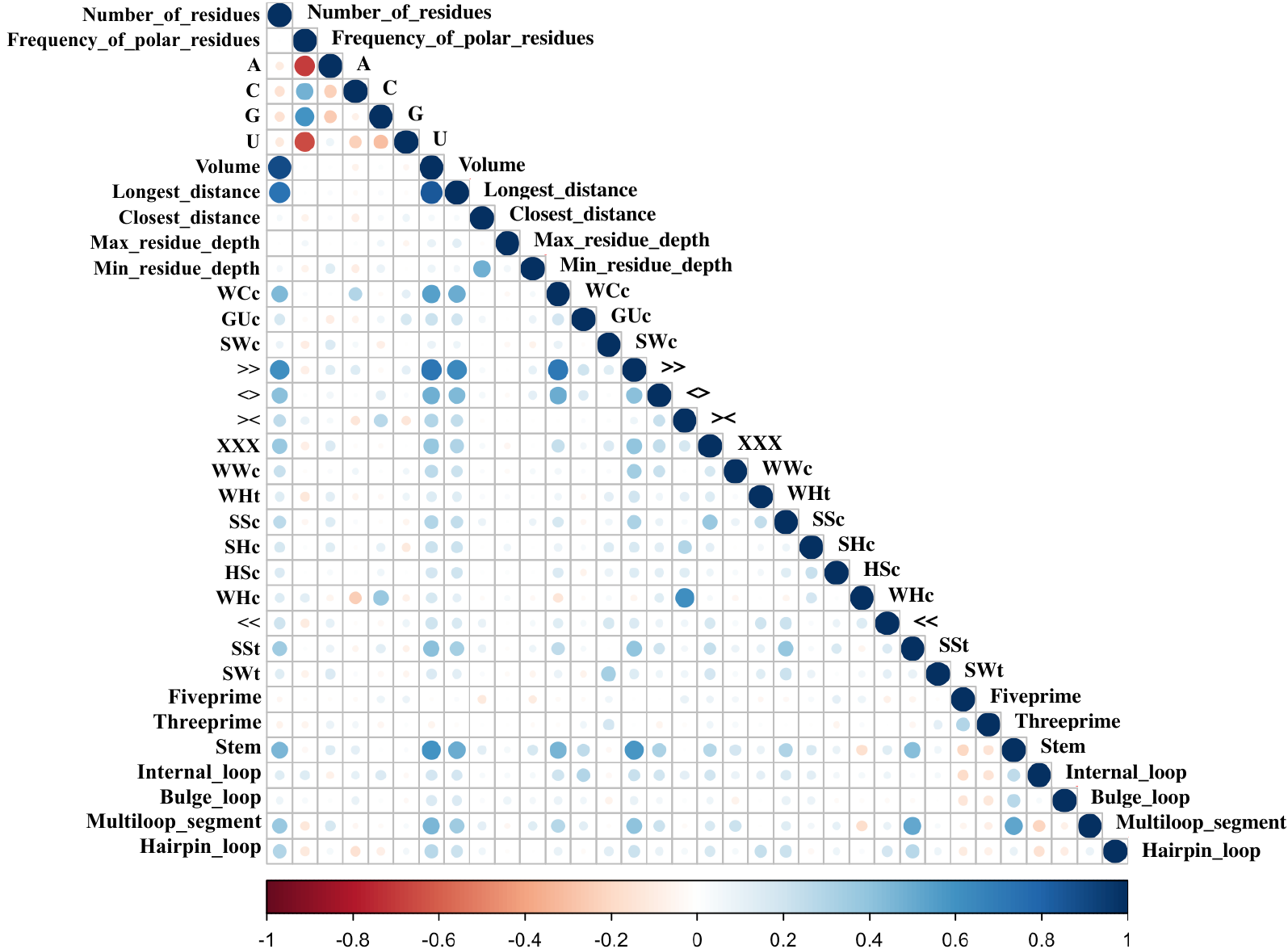}
    \caption{Correlation between 34 descriptors.}
    \label{Fig.corrplot}
\end{figure}

We then build the initial model containing all 34 descriptors. The performance metrics of the initial model are shown in Table~\ref{Tab.performances_model}. The accuracy on the training set is 0.758, which means that approximately 75.8\% of the samples in the training set are correctly classified by the model. The accuracy of test set is slightly lower at 0.706.
In the case of the training set, the sensitivity is calculated to be 0.734, indicating that the model can correctly identify approximately 73.4\% of the approximate binding sites in the training set. For the test set, the sensitivity drops slightly to 0.674.
Regarding specificity, the initial model achieves a specificity of 0.774 on the training set, meaning that it correctly classifies around 77.4\% of the approximate non-binding sites. For the test set, the specificity is slightly lower at 0.737.
The PPV of the model on the training set is 0.767, indicating that 76.7\% of the predicted approximate binding sites are approximate binding sites. For the test set, the PPV drops slightly to 0.719.
The MCC on the training set is calculated to be 0.517, suggesting that the model performs reasonably well in terms of overall predictive ability. The MCC of test set is slightly lower at 0.412.

\begin{table}[h]
    \centering
    \scalebox{0.8}{
    \begin{tabular}{|c|c|c|c|c|c|}
    \hline
    ~ & \multicolumn{2}{|c|}{Initial model}&~& \multicolumn{2}{|c|}{Simplified model}\\
    \cline{1-3}\cline{5-6}
    ~ & \multicolumn{2}{|c|}{34 descriptors}&~& \multicolumn{2}{|c|}{20 descriptors}\\
    \cline{1-3}\cline{5-6}
    Performances&Training set&Test set&~&Training set&Test set\\
    \cline{1-3}\cline{5-6}
    Accuracy&0.758&0.706&~&0.752&0.691\\
    \cline{1-3}\cline{5-6}
    Sensitivity&0.743&0.675&~&0.736&0.664\\
    \cline{1-3}\cline{5-6}
    Specificity&0.774&0.737&~&0.767&0.718\\
    \cline{1-3}\cline{5-6}
    PPV&0.767&0.719&~&0.760&0.702\\
    \cline{1-3}\cline{5-6}
    MCC&0.517&0.412&~&0.504&0.383\\
    \hline
    \multicolumn{4}{|c|}{10-fold cross-validation}&\multicolumn{2}{|c|}{Accuracy: 0.744}\\
    \hline
    \end{tabular}
    }
    \caption{Performance of logistic regression models before and after model simplification using AIC~\cite{Burnham1998} and the p-value associated with the z-value of descriptors~\cite{Becker1988, Johnson1996,Venables2003}.}
    \label{Tab.performances_model}
\end{table}

Next, we apply AIC~\cite{Burnham1998} and the p-value associated with the z-value of descriptors~\cite{Becker1988, Johnson1996,Venables2003} to simplifier the initial model. The initial model includes all 34 descriptors, while the simplified model consists of only 20 descriptors. Table~\ref{Tab.model} presents the estimate, standard error, z-value, and p-value associated with the z-value for each descriptor in the simplified model. All the p-values are less than 0.05, indicating a statistically significant relationship between each predictor descriptor and the response descriptor. Additionally, Table~\ref{Tab.performances_model} also shows the performance of the simplified model. Interestingly, we observe no significant difference in performance between the initial model and the simplified model, both models perform well.

\begin{table}[h]
    \centering
    \scalebox{0.8}{
    \begin{tabular}{|c|c|c|c|c|}
    \hline
    ~& Estimate & Standard error & Z-value &  Pr(>|z|)\\
    \hline
    \hline
    Constant & -0.206 &   0.058  &-3.583 & $0.34 \times 10^{-3}$\\
    Number\_of\_residues&  1.236  &  0.136 &  9.107 & $< 2\times 10^{-16}$\\
A & 0.762  & 0.101 & 7.525 & $5.26\times 10^{-14}$\\
C & 1.116 & 0.096 & 11.676 & $< 2\times 10^{-16}$\\
G & 0.888 & 0.120 & 7.427 & $1.11\times 10^{-13}$\\
U & 0.795 & 0.096 &  8.297 & $< 2\times 10^{-16}$\\
Longest\_distance & 0.215 & 0.093 & 2.309 & 0.02\\  
Closest\_distance & -0.216 & 0.064 & -3.356 & $7.92\times 10^{-4}$\\
Max\_residue\_depth & -0.109 & 0.053 & -2.081 & 0.04\\  
Min\_residue\_depth & -0.186 & 0.062 & -3.003 & $2.67\times 10^{-3}$\\ 
GUc & -0.184 & 0.054 & -3.384 & $7.15\times 10^{-4}$\\
<> & -0.255 & 0.063 & -4.039 & $5.38\times 10^{-5}$\\
WWc & -0.178 & 0.055 & -3.231 & $1.23\times 10^{-3}$\\ 
SHc & 0.196 & 0.057 &  3.444 & $5.73\times 10^{-4}$\\
WHc & 0.989 & 0.120  & 8.252 & $< 2\times 10^{-16}$\\
SWt & -0.154 & 0.058 & -2.663 &$7.75\times 10^{-3}$\\ 
Fiveprime & 0.138 & 0.055 & 2.502 & 0.01\\  
Threeprime & 0.175 & 0.055 & 3.210 & $1.33\times 10^{-3}$\\ 
Bulge\_loop & -0.343 & 0.059 & -5.858 & $4.69\times 10^{-9}$\\
Multiloop\_segment & 0.275 & 0.062 & 4.453 & $8.49\times 10^{-6}$\\
Hairpin\_loop & -0.554 & 0.059 & -9.384 & $< 2\times 10^{-16}$\\
    \hline
    \end{tabular}
    }
    \caption{Summary of 20 descriptors for simplified model (The estimate, standard error, z-value, and p-value associated with the z-value for each descriptor in the simplified model).}
    \label{Tab.model}
\end{table}

\begin{figure}[h]
    \centering
    \includegraphics[width=13cm]{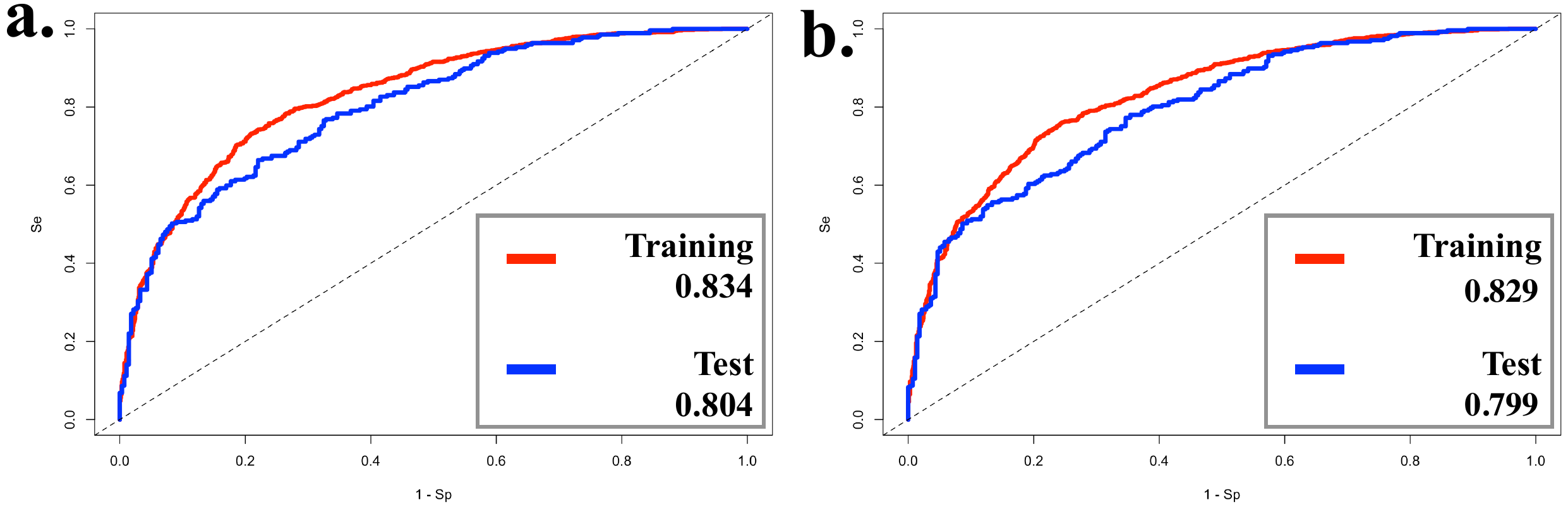}
    \caption{ROC curve of logistic regression models before and after model simplification using AIC~\cite{Burnham1998} and the p-value associated with the z-value of descriptors~\cite{Becker1988, Johnson1996,Venables2003}. a. ROC curve of training set and test set for the initial model. b. ROC curve of training set and test set for the simplified model.}
    \label{Fig.roc}
\end{figure}

Figure~\ref{Fig.roc} shows the ROC curve~\cite{Hanley1982,FAWCETT2006} of each model. Both models exhibit AUC value around 0.83 on the training set and approximately 0.80 on the test set, indicating good performance for both models. There is no significant change observed before and after model simplification. Therefore, the simplified model is chosen as the final model due to its excellent performance and minimum number of descriptors.

\section{Conclusion and discussion}\label{sec4}

We developed \textit{CplxCavity}, a new tool for RNA-small molecule binding sites prediction, by combining a new algorithm and a machine learning model. The algorithm is used to calculate the cavities on the surface based on the 3D structure of the RNA or RNA complex. 
We successfully applied our algorithm to a dataset consisting of 330 RNA-small molecule complexes. The evaluation based on PID revealed that our algorithm accurately calculated 98.07\% binding sites of these complexes. This impressive result indicates that the binding sites are included in the cavities present on the surface of the structure. The algorithm demonstrates its effectiveness in identifying and capturing these binding sites within the surface cavities of the complexes.  Consequently, our algorithm proves to be a reliable and valuable tool for locating binding sites in complex structures.

According to the PID value, we separated the cavities determined by the algorithm into two groups, the approximate binding and non-binding sites. We then used the same number of the approximate binding and non-binding sites, and computed their properties (See Appendix) to create a new dataset for constructing machine learning models. These properties contain geometric, physicochemical descriptors and RNA secondary structure types. The logistic regression model was selected for our study as a binary classification. In the process of constructing the model, we applied AIC~\cite{Burnham1998} and the p-values associated with the z-values of descriptors~\cite{Becker1988, Johnson1996,Venables2003} for model simplification. The final model performs well with only 20 descriptors.

In conclusion, our tool has shown efficiency compared to existing methods. It can be effectively applied in the analysis of RNA-ligand interactions, thereby creating new opportunities to expedite drug discovery and advance the development of RNA-targeted therapeutics. Nevertheless, there is potential for further enhancing the performance of our tools by incorporating additional descriptors, exploring other machine learning models, or integrating an approach based on evolutionary conservation of RNA sequences and structures.

\newpage
\small
\bibliography{bib}
\bibliographystyle{unsrt}

\newpage
\normalsize
\appendix
\section{Appendix}

\subsection{Descriptors for machine learning model}

\begin{itemize}
\item Geometric descriptors
\begin{itemize}
    \item Number of total residues
    \item Volume
    \item Area
    \item Longest distance in the convex hull
    \item Closest distance in the convex hull
    \item Maximum of residue depth (comparing with MSMS (\textit{Maximal speed molecular surface}) surface)~\cite{cock2009}
    \item Minimum of residue depth (comparing with MSMS surface)~\cite{cock2009}
\end{itemize}

\item Physicochemical descriptors
\begin{itemize}
\item  Frequency of polar residues (C, G, CYS, ASP, GLU, HIS, LYS, ARG, SER, THR, TYR, ASN, GLN, ARG, TRP)
\item  Frequency of hydrophobic residues (GLY, ALA, VAL, LEU, ILE, MET, PHE, PRO, TRP)
\item  Frequency of positive residues (ILE, LEU, VAL)
\item  Frequency of negative residues (ASP, GLU)
\item  Frequency of aromatic residues (PHE, TYR, HIS, TRP)
\item  Frequency for each residue (A, C, G, U, T, ALA, ARG, ASN, ASP, CYS, GLU, GLN, GLY, HIS, ILE, LEU, LYS, MET, PHE, PRO, SER, THR, TRP, TYR, VAL)
\item  Stacking ( $>>$:Upward, $<>$:Outward, $<<$:Downward, $><$:Inward )~\cite{Parisien2009, Bottaro2019}
\item  Base pairs (SSt, XXX, WWc, WHc, HWt, WSc, HSc, SSc, WCc, SWc, HHc, SHc, GUc, WHt, WSt, SWt, SHt, HWc, WWt, HHt, HSt)\\
with W = Watson-Crick edge, H = Hoogsteen edge, S= Sugar edge, c/t = cis/trans, XXx = when two bases are in close to each other, but cannot be categorized, this often occurs in low-resolution structures or from molecular simulations. this happens frequently for low-resolution structures or from molecular simulations. WWc pairs between complementary bases are called WCc or GUc.~\cite{Bottaro2019, WATSON1953, Hoogsteen1963}.
\end{itemize}

\item RNA secondary structure types~\cite{Thiel2019}
\begin{itemize}
\item Fiveprime (The unpaired nucleotides at the 5’ end of a molecule/ chain.)
\item Threeprime (The unpaired nucleotides at the 3’ end of a molecule/ chain.)
\item Stem (Regions of contiguous canonical Watson-Crick base-paired nucleotides~\cite{WATSON1953, Hoogsteen1963}.)
\item Internal loop
\item Bulged loop
\item Multiloop segment (Single-stranded regions between two stems, pseudo-knots and exterior loops segments between stems are treated as multiloop segments.)
\item Hairpin loop
\end{itemize}
\end{itemize}
\pagenumbering{gobble}

\end{document}